\documentclass[aps,pra,twocolumn,showpacs,amsfonts,superscriptaddress,floatfix]{revtex4}
\usepackage{pdfsync} 
\usepackage{amsmath}
\usepackage{graphicx}
\usepackage{multimedia}
\usepackage{color}
\usepackage{bbold}
\usepackage{bm}
\usepackage{float}

\newcommand{\beq}{\begin{equation}}
\newcommand{\eeq}{\end{equation}}
\newcommand{\beqarray}{\begin{eqnarray}}
\newcommand{\eeqarray}{\end{eqnarray}}

\begin{document}

\title{Superfluid-Mott transitions and vortices in the Jaynes-Cummings-Hubbard lattices with time reversal symmetry breaking} 
\author{A.L.C. Hayward} 
\affiliation{School of Physics, University of Melbourne, Victoria 3010, Australia}
\author{A.M. Martin}
\affiliation{School of Physics, University of Melbourne, Victoria 3010, Australia}

\date{\today}

\begin{abstract} We investigate the groundstate behaviour of
Jaynes-Cummings-Hubbard lattices in the presence of a synthetic magnetic field, via a
Gutzwiller ansatz.  Specifically, we study the Mott-Superfluid transition,
and the formation of vortex lattices in the superfluid regime. We find a
suppression of the superfluid fraction due to the frustration induced by the
incommensurate magnetic and spacial lattice lengths. We also predict the
formation of triangular vortex lattices inside the superfluid
regime.\end{abstract}


\maketitle

\begin{figure}
\includegraphics[width=\columnwidth]{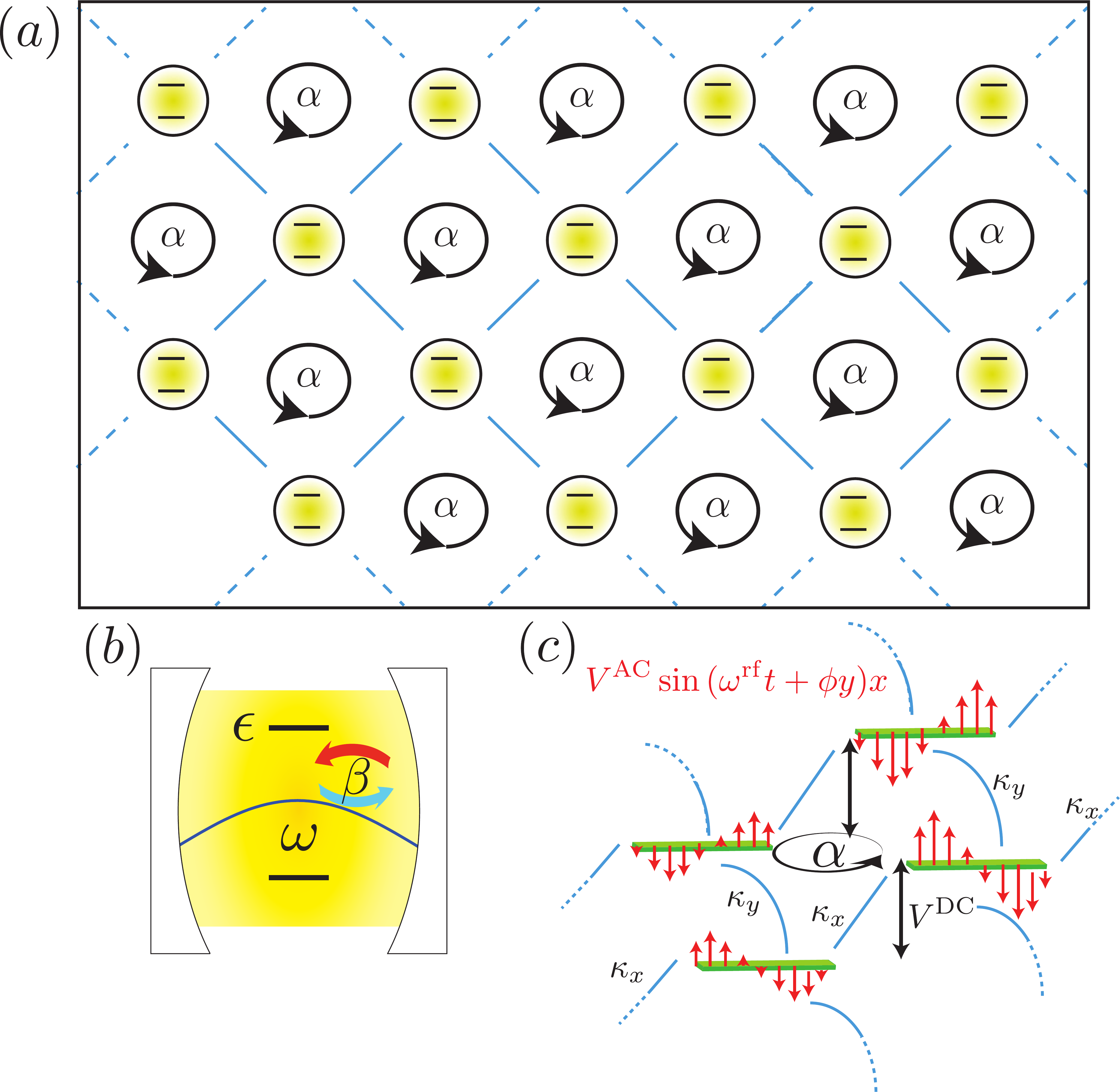}
\vspace{-0.5cm}
\caption{
    (color online). (a) Schematic of a square JCH lattice with a constant synthetic
    magnetic field. Photons moving around a plaquette acquire a phase $2\pi \alpha$
    (b) A single-mode photonic cavity with frequency $\omega$ coupled to a
    two-level atom with strength $\beta$. (c) Scheme for breaking time reversal
    symmetry in photonic cavities: a potential $V=\left[V^{\rm DC} +V^{\rm
    AC}\sin\left(\omega^{\rm rf}+2\pi \alpha y \right)\right]x$, where $x$ and $y$
    are in units of the lattice spacing, is applied to the cavities by dynamically
    tuning $\omega$. The phase offset, $2 \pi \alpha$, along $y$ results in the
    synthetic magnetic field seen in (a).}
\label{fig:lattice}
\vspace{-0.5cm}
\end{figure}

A Jaynes-Cummings Hubbard (JCH) lattice \cite{Hartmann2006, Greentree2006,
Angelakis2007} consists of an array of coupled photonic cavities, with each
cavity mode coupled to a two-level atom [Figs. \ref{fig:lattice} (a) and
\ref{fig:lattice}(b)].  The JCH model is predicted to exhibit a number of solid
state phenomena, including Mott and superfluid phases \cite{Greentree2006},
supersolid phases \cite{Bujnowski2014}, semiconductor physics \cite{Quach2009}
Josephson effect \cite{Gerace2009}, metamaterial properties \cite{Quach2011a},
and Bose-glass phases \cite{Kapit2014} and, in the presence of synthetic magnetic
fields, fractional quantum Hall states \cite{Rossini2007}. 

In this paper we examine how the introduction of a synthetic magnetic field
affects the superfluid Mott insulator phase transition and leads to the
formation of vortices in the superfluid phase. In section \ref{sec:JCHM} we
introduce the formalism and numerical techniques under which we will study the
JCH model. Section \ref{sec:PhaseDiagram} examines the properties of the
phase transition, and how the presence of a synthetic magnetic field affects the
Mott-superfluid transition. Our results suggest that frustration due to
incommensurate magnetic/spacial lattices suppresses the emergence of a
superfluid phase at strong non-linear interaction strengths. Finally, in
section \ref{sec:Vortex}, we study the formation of vortices in the photonic
field in the superfluid regime. We find that vortices in the photonic field of
the JCH behave in a similar way to those found in continuous superfluids,
including the formation of a triangular Abrikosov lattice.

\section{\label{sec:JCHM} Jaynes-Cummings-Hubbard Model}
Each cavity [Fig. \ref{fig:lattice}(b)] in the JCH lattice is described by the Jaynes-Cummings (JC) Hamiltonian ($\hbar =1$)
\begin{eqnarray}
H^{\rm JC}=\omega L+\Delta \sigma^+ \sigma^- +\beta\left( \sigma^{+} a +\sigma^- a^{\dagger}\right),
\label{eq:HJC}
\end{eqnarray}
where $a$ is the photonic annihilation operator, $\sigma^{\pm}$ are the atom
raising and lowering operators, $\Delta$ is the atom-photon detuning and
$\beta$ is the coupling energy. For $n$ photons the state $| g (e),n \rangle$
forms the single cavity basis, where $g (e)$ denote the ground (excited) state
of the atom. Throughout this paper we shall be using the dressed basis 
\begin{eqnarray}
\label{equ:JCHsol}
| \pm ,\ell \rangle=\frac{\beta \sqrt{\ell}|g,\ell\rangle +\left[-\Delta/2 \pm \chi\right]|e,\ell-1\rangle}{\sqrt{2\xi^2\mp\xi\Delta}}, 
\end{eqnarray}
with 
\begin{eqnarray}
\chi=\sqrt{\ell\beta^2 +\Delta^2/4}
\end{eqnarray}
the generalized Rabi frequency, and
$\ell$ refers to the total number of excitations in the cavity.  In this basis the energy of a single cavity, with $\ell$ excitations is
\begin{eqnarray}
E_{\pm,\ell}=\ell\omega \pm \chi -\Delta/2.
\end{eqnarray}

Since $H^{\rm JC}$ commutes with the total excitation number operator, $L
=a^{\dagger}a +\sigma^{+}\sigma^-$ the total excitations in the cavity, $\ell$, is
a good quantum number. The eigenstates of Eq.~(\ref{eq:HJC}) are termed
polaritons, superpositions of atomic and photonic excitations, and are a
function of $\ell$ and $\Delta/\beta$.

The JCH model describes a tight-binding JC lattice [Fig. \ref{fig:lattice}(a) with $\alpha=0$]:
\begin{eqnarray}
H^{\rm JCH}= H^{\rm JC}+K=\sum_i^N H_i^{\rm JC} - \sum_{\langle i,j \rangle} \kappa_{ij} a^{\dagger}_i a_j,
\end{eqnarray}
where $\kappa_{ij}$ is the photon tunnelling rate between cavities $i$ and $j$ and the sum over $\langle i,j\rangle$ is between nearest neighbours only.

Exact solutions to $H^{\rm JCH }$ are know in two limits; $\kappa/\beta \rightarrow
0$, and $\beta/\kappa \rightarrow 0$. In the former limit, each cavity is decoupled,
and the eigenstates are simply product states of individual cavities,
described by the solutions in Eq.~(\ref{equ:JCHsol}). In the opposite limit,
where $\beta/\kappa \rightarrow 0$, the photonic and atomic degrees of freedom decouple. This
removes the non-linearity from the photonic field, and any inter-atomic
coupling, again reducing the eigenstates to product states of single-particle. 

\subsection{The Gutzwiller Ansatz}
For large system sizes, the exact  quantum mechanical problem of the JCH model
is too computationally difficult to approach directly, either analytically or
numerically. Various  approximations have been used to study the superfluid Mott insulator phase transition in the JCH model \cite{Greentree2006,Hohenadler2011}.
In this paper we use a Gutzwiller ansatz to study the JCH groundstate
wavefunction. The Gutzwiller ansatz imposes a trial wavefunction that is a
product state of individual lattice sites:
\begin{equation}
	 P_\ell|\Psi^{\rm GW}\rangle = \prod\limits_i^{\rm sites} |\psi_i\rangle
\end{equation}
where $P_\ell$ projects onto an excitation number eigenstate. Applying 
$|\Psi^{\rm GW}\rangle$ to the JCH Hamiltonian yields an effective Hamiltonian that
is a sum over single cavity Hamiltonians:
\begin{eqnarray}
H^{\rm JCH}_{\rm eff}= \sum_i^N H_i^{\rm JC} - \sum_{\langle i,j \rangle} \left[\kappa_{ij} \psi_j^{\star}\left( a_i-\frac{1}{2}\psi_i\right) +{\rm h.c.}\right],
\label{eq:JCH_eff}
\end{eqnarray}
where we have introduced the notation $\psi_i=\langle \psi_i | a_i |\psi_i
\rangle$.  With this
effective Hamiltonian in hand, the Gutzwiller groundstate can be found by
minimising the energy of $H^{\rm JCH}_{\rm eff}$ .

The local superfluid order parameter $\vec{\psi} = \{\psi_i\}$ determines the
nature of the groundstate. From Eq.~(\ref{eq:JCH_eff}) it follows that the off
diagonal elements of the single particle density matrix are determined by
$\vec{\psi}$: $\rho_{ij}^1=\psi_i \psi_j^{\star}$.  The system can be said to be
superfluid whenever there are system wide connected non-zero $\psi_i$'s. The
Mott insulating phase is characterised by $\psi_i = 0$ across the whole system.

The effective Hamiltonian arrived at though the Gutzwiller procedure removes
the restriction of the system to a specific number of excitations, since,
except in the $\vec{\psi}=0$ case, the Hamiltonian does not conserve the particle
number. However, it is possible to restrict the particle density implicitly
though the introduction of a chemical potential:
\begin{eqnarray}
\label{equ:muHam}    
H=H^{\rm
JCH}-\mu L.
\end{eqnarray}
That is, the system is now studied in the
grand-canonical ensemble, and the system's groundstate at the desired density
is found by varying the chemical potential.  In the case of the effective
Hamiltonian in Eq.~(\ref{eq:JCH_eff}), this is necessary, since this
Hamiltonian does not preserve photon number.

There has been some confusion in the literature\cite{Greentree2006,Zheng2011}
as to the physical
significance, or meaning, of the chemical potential in photonic systems. From a
mathematical perspective, the chemical potential introduced in this manner acts
simply as a Lagrange multiplier. The choice of $\mu$ defines an excitation
density for the groundstate. The chemical potential can also appear as a real
physical quantity. Consider the JCH weakly coupled to a bath of photons, all
with frequency $\omega_b$ such that $\omega-\omega_b=\mu$. Assuming a mechanism
for thermalization, such a coupling of the bath to atomic or photonic modes,
the pump will act to drive the system to some groundstate, determined by
$\mu=\omega_{pump}$ . Weakly blue band pumping will act to create a positive chemical
potential for photons in the system. Although a JCH system under such pumping,
when paired with photon losses, will exhibit non-equilibrium dynamics, when
these effects are small, the system will be well described by Eq. (\ref{equ:muHam}) in the groundstate\cite{Hu2013}. 

For real world cavity lattices, the photons in the system will not be
conserved, due to dissipation. In contrast to cold atom systems, where some set
number of particles can be loaded into the lattice, a photonic lattice is
driven by a laser, and will be in a steady state where losses and driving
equilibrate\cite{Carusotto2013}. Still, for an array of high quality cavities,
one may consider a regime of weak pumping, and weak losses, such that the
mean-field approximation made here captures the broad physics of the
system\cite{LeBoite2014}. 

\subsection{Synthetic Magnetic Field} An artificial magnetic field may  be
realized via the introduction of some time reversal symmetry breaking
interaction. A number of techniques have been proposed to achieve
this\cite{Koch2010a,Kolovsky2011,Umucalilar2011}. For example, one may exploit
a time-dependent potential to induce magnetic flux across the lattice (as in
Fig \ref{fig:lattice}(c))\cite{Kolovsky2011}.

Defining the magnetic field by a vector potential ${\bf A}(x)$ results in
minimal substitution, ${\bf p} \rightarrow {\bf p}-q {\bf A}(x)$. On a
tight-binding lattice, a vector potential ${\bf A}$ manifests as a complex
hopping rate $\kappa_{ij} \rightarrow \kappa_{ij}e^{i2\pi \theta_{ij}}$, where
$2 \pi \theta_{ij}=\int_{r_i}^{r_j}{\bf A}(r) dr$. Gauge symmetry implies that
the only physically important parameter is the total phase, $2\pi \alpha$,
picked up around a closed loop, where $\alpha=\Phi/\Phi_0$ is the fraction of
flux quanta through the loop. A uniform magnetic field through the two
dimensional lattice corresponds to a constant $\alpha$ for all plaquettes on
the lattice. Factors of $2\pi$ in the phase around a loop are physically
inconsequential, so we only need consider $\alpha \in [0,1)$.

\section{ \label{sec:PhaseDiagram}Phase diagram}
\begin{figure}
\includegraphics[width=\columnwidth]{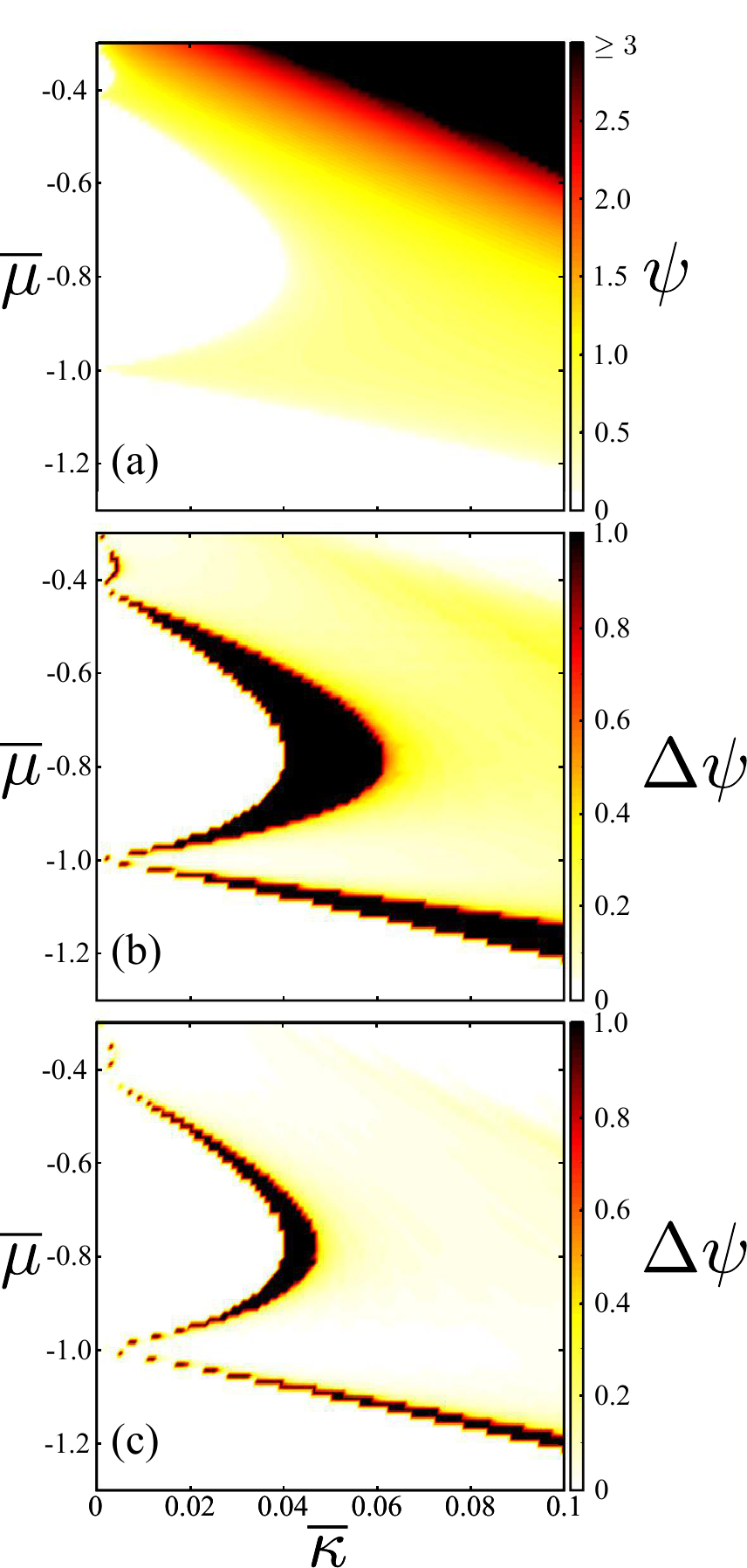}
\caption{
    (color online). (a) The superfluid order parameter ($\psi$) as a function of
    the chemical potential ${\overline \mu}$ and the interactive hopping
    ${\overline \kappa}$, with a lattice of
    $4 \times 5$ sites in the absence of synthetic magnetic field ($\alpha=0$).
    (b,c) $\Delta\psi$ as a
    function of the chemical potential ${\overline \mu}$ and the interactive
    hopping ${\overline \kappa}$, for $|\alpha|=2/5$ and $3/5$. (b) and $|\alpha|=1/10$ and $9/10$.
    (c), with $\epsilon=0.001$. For (a-c) we have introduced the
    following dimensionless parameterisation: ${\overline \kappa}=\kappa/\beta$,
    ${\overline \mu}=(\mu-\omega)/\beta$ and ${\overline \Delta} =
    \Delta/\beta=0$}
\label{fig:mottlobes}
\end{figure}

The simplest situation in which the Mott-superfluid transition occurs is in the
homogeneous lattice, with $\alpha=0$. Since the meanfield ground state of our
Hamiltonian is invariant under translations along lattice vectors it is
possible to  significantly simplify the problem to that of a single site. That
is, since $|\psi_i\rangle = |\psi_j\rangle \forall i,j$ we have just a single
$\psi=\langle a_i \rangle$ that characterises the whole groundstate. Hence for
$\alpha=0$  we find the well-known result\cite{Hartmann2006, Greentree2006,
Angelakis2007} [Fig. \ref{fig:mottlobes}(a)] that
the parameter space is separated into two distinct phases. For low hopping
strength ${\overline \kappa}=\kappa/\beta$, we find lobes of vanishing
superfluid order parameter, i.e., Mott-insulating phases as shown in Fig.
\ref{fig:mottlobes}(a). Each lobe corresponds to a state with an integer number
of strongly localised excitations per site. For example the first (second) Mott
lobe, with ${\overline \mu}=(\mu-\omega)/\beta<-1$ ($-1<{\overline
\mu}=(\mu-\omega)/\beta<-0.4$) corresponds to the states $|-,0\rangle$
($|-,1\rangle$). For low chemical potential, there are no excitations in the
system. Raising the chemical potential, the block is successively filled with
one, two, and more excitations per site. At sufficiently large ${\overline
\kappa}$, the system undergoes a phase transition into a phase of finite
superfluid order parameter [Fig. \ref{fig:mottlobes}(a)]. The excitations are
homogeneously distributed and delocalised, i.e., the system is in a superfluid
state. 

The Mott-superfluid transition in the JCH lattice is now considered in the
presence of a synthetic magnetic field $\alpha \ne 0$. This situation differs
in several key ways from the non-magnetic case. Specifically, the field
introduces a {\it magnetic lattice} which breaks the translational symmetry of
the underlying lattice. Hence, with a magnetic field, $\psi_i$'s can not be
assumed identical across the system. To capture this, a large lattice must be
simulated, so that the magnetic and cavity lattices are commensurate. For the
magnetic parameter $\alpha = p/q$ , there must be $q = N_x \times N_y$ sites,
where $N_x$ ($N_y$) is the number of cavities along the $x$ ($y$) axis.
Despite this introducing a synthetic magnetic field does not qualitatively
change the phase diagram for the Mott-insulator superfluid phase transition,
i.e. at low ${\overline \kappa}$ the lobe structure persists and when
${\overline \kappa}$ is sufficiently large the system undergoes a transition to
superfluid state. However, the introduction of a synthetic magnetic field does
quantitatively change the phase diagram, see Figs. \ref{fig:mottlobes}(b,c).
Here, we compare the superfluid density at non-zero $\alpha$ to the $\alpha=0$
case (at some $\mu$ and $\kappa$)via 
\begin{equation}
    \Delta\psi(\alpha) = \frac{\max\lbrack\vec{\psi}(\alpha=0)] - \max\lbrack\vec{\psi}(\alpha)]}{\max\lbrack\vec{\psi}(\alpha=0)\rbrack + \max\lbrack\vec{\psi}(\alpha)\rbrack+\epsilon},
\end{equation}
where we have introduced $\epsilon$ to avoid divide by zero errors, but which plays no physical role.
Specifically we see that the introduction of a synthetic magnetic field changes
(for a fixed chemical potential) the strength of inter-cavity hopping required
to transition from the Mott-phase to the superfluid phase.
\begin{figure}
\includegraphics[width=\columnwidth]{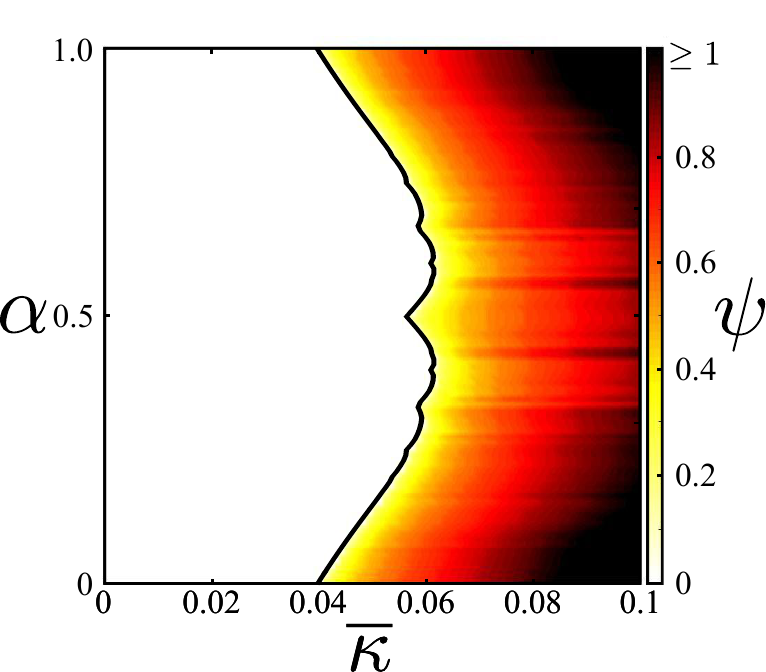}
\caption{(color online). The maximum superfluid order parameter ($\max\lbrack
    \vec{\psi}\rbrack$) as a function of the synthetic magnetic field $\alpha$
    and the interactive hopping ${\overline \kappa}$, with ${\overline
    \mu}=-0.78$ for a $20 \times 20$ lattice. The solid black curve corresponds
    to the evaluation of the transition between the Mott insulator state and
    the superfluid state as determined by Eq.~(\ref{equ:exact_transition}). We
    have introduced the following dimensionless parameterisation: ${\overline
        \kappa}=\kappa/\beta$, ${\overline \mu}=(\mu-\omega)/\beta$ and
    ${\overline \Delta} = (\omega-\epsilon)/\beta=0$}
\label{fig:transition}
\end{figure}

To quantify how the synthetic magnetic field changes the boundary between the
superfluid and Mott phase in Fig. \ref{fig:mottlobes} we have plotted, in Fig. \ref{fig:transition},  $\max\lbrack \vec{\psi}\rbrack$ as a function of ${\overline \kappa}$ and $\alpha$,
with ${\overline \mu}=-0.78$. This shows that increasing the synthetic magnetic
field from zero (half a) flux quanta per site  to half a (one) flux quanta per
site the superfluid Mott insulator boundary is pushed to larger (lower) values
of the inter-cavity hopping.

In the vicinity of the Mott insulator superfluid phase boundary the superfluid
parameters $\psi_i$ are very small across the entire system. Perturbatively
expanding about $\psi_i$ it is possible to determine the boundary between the
superfluid and Mott phases. Specifically, consider the single site effective
Hamiltonian for site $i$
\begin{eqnarray}
H^{\rm JCH}_{{\rm eff},i}=H_i^{\rm JC} -\mu L_i - \kappa \sum_{j} \left[\left( a_i^{\dagger}-\frac{1}{2}\psi_i^{\star}\right)\Psi_i +{\rm h.c.}\right],
\end{eqnarray}
where $\Psi_i=\sum_j \kappa_{ij}\psi_j/\kappa$ and we have assumed that the
magnitude of the hopping rate is the same between sites, i.e.
$|\kappa_{ij}|=\kappa$. Treating the $\psi_i$'s as perturbative parameters we
find
\begin{eqnarray}
    \label{equ:energy_functional}
E_i({\vec \psi})&=&E^{(0)}_i - r_n \kappa^2 \left|\Psi_i\right|^2-\frac{\kappa}{2}\left(\psi_i^{\star} \Psi_i +{\rm h.c.}\right),
\end{eqnarray}  
where 
\begin{eqnarray}
r_n=\sum_{\gamma=\pm} \frac{\left|\langle \gamma, n+1 \left| a\right| -, n\rangle \right|^2}{E_{n+1,\gamma}-E_{n,-}-\mu}+\frac{\left|\langle \gamma, n-1 \left| a\right| -,n\rangle \right|^2}{E_{n-1,\gamma}-E_{n,-}-\mu} \nonumber \\
\end{eqnarray}
arises from the second order perturbative corrections to the energy. The full energy can then be written as
\begin{equation}
E({\vec \psi})=\sum_i E_i^{(0)} +\kappa {\vec
\psi}^{\dagger} {\bm \kappa}(\alpha){\vec \psi}- r_n \kappa^2
{\vec \psi}^{\dagger} {\bm \kappa}^2(\alpha){\vec \psi},
\end{equation}
where ${\vec \psi}$ is a vector of the $\psi_i$'s and the elements
of the matrix ${\bm \kappa}$ parameterise the phase accumulated, due to the
synthetic magnetic field, in hopping between sites. The location of the Mott
insulator superfluid transition occurs when some non-zero values for $\psi_i$
gives an energy lower than the energy of the Mott lobe, i.e.
\begin{eqnarray}
    \label{equ:exact_transition}
\kappa{\vec \psi}^{\dagger} {\bm \kappa}{\vec \psi}- r_n \kappa^2 {\vec \psi}^{\dagger} {\bm \kappa}^2{\vec \psi}< 0.
\end{eqnarray}
The solid line in Fig.~\ref{fig:transition} shows how the phase boundary between the Mott
insulator and superfluid changes as a function of the hopping and $\alpha$, as
determined from the above equation. As expected, we find very good agreement
between this perturbative calculation and the full mean-field calculation in
determining the boundary between the two regimes. 

The solutions for Eq.(\ref{equ:exact_transition}) occur for $\kappa > 1/\kappa f(\alpha)$, where
$f(\alpha)$ is the maximum eigenvalue of ${\bm \kappa}(\alpha)$.  As noted in
studies of the Bose-Hubbard model\cite{Oktel2007}, $f(\alpha)$ is the outline
of the Hofstadter butterfly. Thus the suppression of the phase transition comes
from the frustration induced by the incommensurate lattice/magnetic lengths.
This leads in general to fractal structures (as seen in the Hofstadter
Butterfly), and is similarly reflected in the phase boundary seen here. 

In Fig.~\ref{fig:transition} we note that the some of the very fine structure
in $f(\alpha)$ is visible, both in the location of the boundary, and within the
superfluid region. This structure is perhaps over represented in the figure
since, in finding the groundstate wavefunction, we limited the system size to a
$20 \times 20$ lattice.  The presence of butterfly like structure has already been observed
in experiments\cite{Dean2013,Schlosser1996}, including some involving
photons\cite{Kuhl1998}. As expected\cite{Hofstadter1976a}, small fluctuations
in the system lead to a blurring of the fractal structure, and so the very fine
structure seen in Fig.~\ref{fig:transition} would be challenging to resolve in
any experiment. 

\section{Vortices \label{sec:Vortex}} 

\begin{figure} \includegraphics[width=\columnwidth]{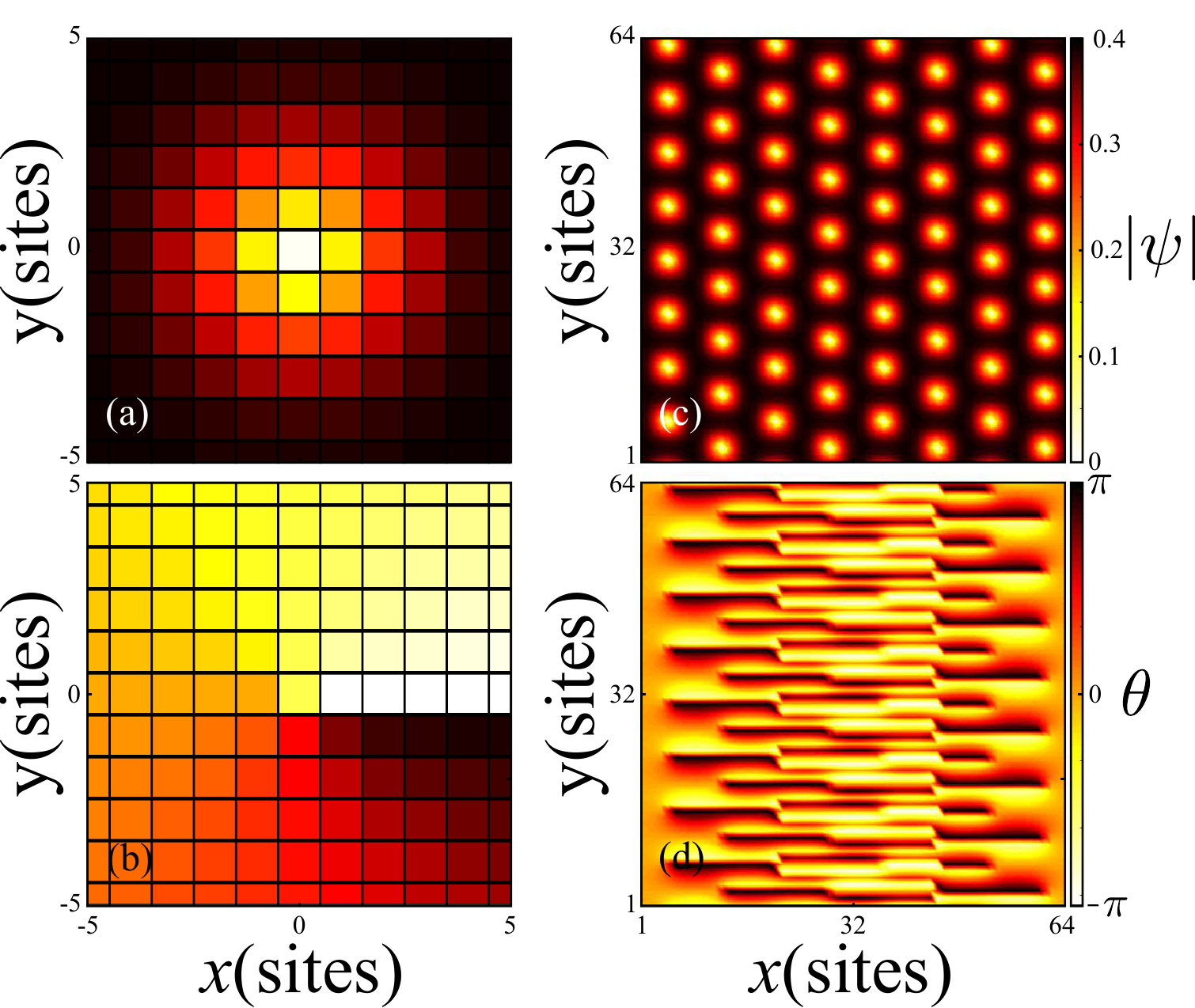}
\caption{ (color online). Superfluid density $|\psi|$ (a,c) and phase (b,d) on
    a $64 \times 64$ lattice, with  ${\overline \kappa}=0.045$, ${\overline \Delta}=0$ and ${\overline
        \mu}=-0.78$ for $\alpha=1/64^2$ (a,b) and $\alpha=1/64$ (c,d).} \label{fig:vorticies}
\end{figure}

In dilute Bose gas superfluids the introduction of synthetic magnetic
fields\cite{Ruostekoski2002,Jaksch2003,Mueller2004,Spielman2009,Murray2009}, or
the application of rotation to break time reversal symmetry, has lead to the
observation of single vortices\cite{Lin2009}, vortex
lattices\cite{Madison2000,Madison2001,Hodby2001}, and the prediction of the
emergence of fractional quantum Hall
states\cite{Cooper1999,Wilkin2000,Cooper2001,Ruostekoski2002,Bretin2004,Schweikhard2004,Baranov2005}.
The application of a synthetic magnetic field in a JCH lattice is expected to
produce similar effects in the superfluid state. Considerable effort has
already been applied to the study of of fractional quantum Hall
states\cite{Kapit2014} in the `high' synthetic magnetic field regime, where the
number of flux quanta though the lattice is larger than the number of
excitations.  However, less attention has been paid to the emergence of a
vortex and vortex lattice states in the `low' synthetic magnetic field regime.
Below we demonstrate that the JCH system does admit vortex and vortex lattice
solutions in the superfluid regime, upon the introduction of a synthetic
magnetic field. 

In the context of the meanfield description of JCH model, the local order
parameter shares all the characteristics of a superfluid. Figures
\ref{fig:vorticies}(a,b) show the magnitude of the mean field order parameter
(a) and its phase (b) for a periodic lattice with a single flux quantum
penetrating $64 \times 64$ sites. Minimizing the meanfield energy
functional(Eq.~\ref{equ:energy_functional}) results in a single vortex
structure where the superfluid density rises monotonically from the centre of
the vortex core and the superfluid phase rotates by $2\pi$. 

At higher synthetic magnetic fields one expects multiple vortices to be
admitted into the superfluid. In the absence of an underlying lattice structure
the vortices will arrange themselves into a triangular Abrikosov vortex
lattice. This configuration arises, in the presence of a local repulsive
non-linearity, to minimize the self-interaction of the superfluid at a
given density of vortices. Figures \ref{fig:vorticies}(c,d) show that energy
minimization of the meanfield JCH, in the presence of a synthetic magnetic
field, also leads to the formation of an Abrikosov vortex lattice state. 

\section{Conclusions} We have shown that the introduction of a
synthetic magnetic field to the JCH model modifies the boundary between the
Mott-Insulator and superfluid regimes. This modification arises from a
competition between the magnetic lattice and the spatial lattice. Additionally,
we predict that in the superfluid regime the introduction of a synthetic
magnetic field leads to the formation of vortices which, due to the local non-linear
atom-photon interaction, forms a triangular lattice in the groundstate. 

This work opens up possible avenues for future research into the role of the
JCH atom-cavity interaction in frustrating the formation of triangular
Abrikosov vortex lattices, and, since this is an inherently two dimensional
system, the possibility of observing a BKT transition.
\bibliographystyle{apsrev}
\bibliography{library}

\begin{thebibliography}{37}
\expandafter\ifx\csname natexlab\endcsname\relax\def\natexlab#1{#1}\fi
\expandafter\ifx\csname bibnamefont\endcsname\relax
  \def\bibnamefont#1{#1}\fi
\expandafter\ifx\csname bibfnamefont\endcsname\relax
  \def\bibfnamefont#1{#1}\fi
\expandafter\ifx\csname citenamefont\endcsname\relax
  \def\citenamefont#1{#1}\fi
\expandafter\ifx\csname url\endcsname\relax
  \def\url#1{\texttt{#1}}\fi
\expandafter\ifx\csname urlprefix\endcsname\relax\def\urlprefix{URL }\fi
\providecommand{\bibinfo}[2]{#2}
\providecommand{\eprint}[2][]{\url{#2}}

\bibitem[{\citenamefont{Hartmann et~al.}(2006)\citenamefont{Hartmann,
  Brand\~{a}o, and Plenio}}]{Hartmann2006}
\bibinfo{author}{\bibfnamefont{M.~J.} \bibnamefont{Hartmann}},
  \bibinfo{author}{\bibfnamefont{F.~G. S.~L.} \bibnamefont{Brand\~{a}o}},
  \bibnamefont{and} \bibinfo{author}{\bibfnamefont{M.~B.}
  \bibnamefont{Plenio}}, \bibinfo{journal}{Nature Physics}
  \textbf{\bibinfo{volume}{2}}, \bibinfo{pages}{849} (\bibinfo{year}{2006}),
  ISSN \bibinfo{issn}{1745-2473}.

\bibitem[{\citenamefont{Greentree et~al.}(2006)\citenamefont{Greentree, Tahan,
  Cole, and Hollenberg}}]{Greentree2006}
\bibinfo{author}{\bibfnamefont{A.~D.} \bibnamefont{Greentree}},
  \bibinfo{author}{\bibfnamefont{C.}~\bibnamefont{Tahan}},
  \bibinfo{author}{\bibfnamefont{J.~H.} \bibnamefont{Cole}}, \bibnamefont{and}
  \bibinfo{author}{\bibfnamefont{L.~C.~L.} \bibnamefont{Hollenberg}},
  \bibinfo{journal}{Nature Physics} \textbf{\bibinfo{volume}{2}},
  \bibinfo{pages}{856} (\bibinfo{year}{2006}), ISSN \bibinfo{issn}{1745-2473}.

\bibitem[{\citenamefont{Angelakis et~al.}(2007)\citenamefont{Angelakis, Santos,
  and Bose}}]{Angelakis2007}
\bibinfo{author}{\bibfnamefont{D.}~\bibnamefont{Angelakis}},
  \bibinfo{author}{\bibfnamefont{M.}~\bibnamefont{Santos}}, \bibnamefont{and}
  \bibinfo{author}{\bibfnamefont{S.}~\bibnamefont{Bose}},
  \bibinfo{journal}{Physical Review A} \textbf{\bibinfo{volume}{76}}
  (\bibinfo{year}{2007}), ISSN \bibinfo{issn}{1050-2947}.

\bibitem[{\citenamefont{Bujnowski et~al.}(2014)\citenamefont{Bujnowski, Corso,
  Hayward, Cole, and Martin}}]{Bujnowski2014}
\bibinfo{author}{\bibfnamefont{B.}~\bibnamefont{Bujnowski}},
  \bibinfo{author}{\bibfnamefont{J.~K.} \bibnamefont{Corso}},
  \bibinfo{author}{\bibfnamefont{A.~L.~C.} \bibnamefont{Hayward}},
  \bibinfo{author}{\bibfnamefont{J.~H.} \bibnamefont{Cole}}, \bibnamefont{and}
  \bibinfo{author}{\bibfnamefont{A.~M.} \bibnamefont{Martin}},
  \bibinfo{journal}{Physical Review A} \textbf{\bibinfo{volume}{90}},
  \bibinfo{pages}{043801} (\bibinfo{year}{2014}), ISSN
  \bibinfo{issn}{1050-2947}.

\bibitem[{\citenamefont{Quach et~al.}(2009)\citenamefont{Quach, Makin, Su,
  Greentree, and Hollenberg}}]{Quach2009}
\bibinfo{author}{\bibfnamefont{J.}~\bibnamefont{Quach}},
  \bibinfo{author}{\bibfnamefont{M.~I.} \bibnamefont{Makin}},
  \bibinfo{author}{\bibfnamefont{C.-H.} \bibnamefont{Su}},
  \bibinfo{author}{\bibfnamefont{A.~D.} \bibnamefont{Greentree}},
  \bibnamefont{and} \bibinfo{author}{\bibfnamefont{L.~C.~L.}
  \bibnamefont{Hollenberg}}, \bibinfo{journal}{Physical Review A}
  \textbf{\bibinfo{volume}{80}} (\bibinfo{year}{2009}), ISSN
  \bibinfo{issn}{1050-2947}.

\bibitem[{\citenamefont{Gerace et~al.}(2009)\citenamefont{Gerace, T\"{u}reci,
  Imamoglu, Giovannetti, and Fazio}}]{Gerace2009}
\bibinfo{author}{\bibfnamefont{D.}~\bibnamefont{Gerace}},
  \bibinfo{author}{\bibfnamefont{H.~E.} \bibnamefont{T\"{u}reci}},
  \bibinfo{author}{\bibfnamefont{A.}~\bibnamefont{Imamoglu}},
  \bibinfo{author}{\bibfnamefont{V.}~\bibnamefont{Giovannetti}},
  \bibnamefont{and} \bibinfo{author}{\bibfnamefont{R.}~\bibnamefont{Fazio}},
  \bibinfo{journal}{Nature Physics} \textbf{\bibinfo{volume}{5}},
  \bibinfo{pages}{281} (\bibinfo{year}{2009}), ISSN \bibinfo{issn}{1745-2473}.

\bibitem[{\citenamefont{Quach et~al.}(2011)\citenamefont{Quach, Su, Martin,
  Greentree, and Hollenberg}}]{Quach2011a}
\bibinfo{author}{\bibfnamefont{J.~Q.} \bibnamefont{Quach}},
  \bibinfo{author}{\bibfnamefont{C.-H.} \bibnamefont{Su}},
  \bibinfo{author}{\bibfnamefont{A.~M.} \bibnamefont{Martin}},
  \bibinfo{author}{\bibfnamefont{A.~D.} \bibnamefont{Greentree}},
  \bibnamefont{and} \bibinfo{author}{\bibfnamefont{L.~C.~L.}
  \bibnamefont{Hollenberg}}, \bibinfo{journal}{Optics express}
  \textbf{\bibinfo{volume}{19}}, \bibinfo{pages}{11018} (\bibinfo{year}{2011}),
  ISSN \bibinfo{issn}{1094-4087}.

\bibitem[{\citenamefont{Kapit et~al.}(2014)\citenamefont{Kapit, Hafezi, and
  Simon}}]{Kapit2014}
\bibinfo{author}{\bibfnamefont{E.}~\bibnamefont{Kapit}},
  \bibinfo{author}{\bibfnamefont{M.}~\bibnamefont{Hafezi}}, \bibnamefont{and}
  \bibinfo{author}{\bibfnamefont{S.~H.} \bibnamefont{Simon}},
  \bibinfo{journal}{Physical Review X} \textbf{\bibinfo{volume}{4}},
  \bibinfo{pages}{1} (\bibinfo{year}{2014}), ISSN \bibinfo{issn}{2160-3308},
  \eprint{arXiv:1402.6847v1}.

\bibitem[{\citenamefont{Rossini and Fazio}(2007)}]{Rossini2007}
\bibinfo{author}{\bibfnamefont{D.}~\bibnamefont{Rossini}} \bibnamefont{and}
  \bibinfo{author}{\bibfnamefont{R.}~\bibnamefont{Fazio}},
  \bibinfo{journal}{Physical Review Letters} \textbf{\bibinfo{volume}{99}}
  (\bibinfo{year}{2007}), ISSN \bibinfo{issn}{0031-9007}.

\bibitem[{\citenamefont{Hohenadler et~al.}(2011)\citenamefont{Hohenadler,
  Aichhorn, Schmidt, and Pollet}}]{Hohenadler2011}
\bibinfo{author}{\bibfnamefont{M.}~\bibnamefont{Hohenadler}},
  \bibinfo{author}{\bibfnamefont{M.}~\bibnamefont{Aichhorn}},
  \bibinfo{author}{\bibfnamefont{S.}~\bibnamefont{Schmidt}}, \bibnamefont{and}
  \bibinfo{author}{\bibfnamefont{L.}~\bibnamefont{Pollet}},
  \bibinfo{journal}{Physical Review A} \textbf{\bibinfo{volume}{84}}
  (\bibinfo{year}{2011}), ISSN \bibinfo{issn}{1050-2947}.

\bibitem[{\citenamefont{Zheng and Takada}(2011)}]{Zheng2011}
\bibinfo{author}{\bibfnamefont{H.}~\bibnamefont{Zheng}} \bibnamefont{and}
  \bibinfo{author}{\bibfnamefont{Y.}~\bibnamefont{Takada}},
  \bibinfo{journal}{Physical Review A} \textbf{\bibinfo{volume}{84}},
  \bibinfo{pages}{043819} (\bibinfo{year}{2011}), ISSN
  \bibinfo{issn}{1050-2947}.

\bibitem[{\citenamefont{Hu et~al.}(2013)\citenamefont{Hu, Lee, and
  Clark}}]{Hu2013}
\bibinfo{author}{\bibfnamefont{A.}~\bibnamefont{Hu}},
  \bibinfo{author}{\bibfnamefont{T.~E.} \bibnamefont{Lee}}, \bibnamefont{and}
  \bibinfo{author}{\bibfnamefont{C.~W.} \bibnamefont{Clark}},
  \bibinfo{journal}{Physical Review A} \textbf{\bibinfo{volume}{88}},
  \bibinfo{pages}{053627} (\bibinfo{year}{2013}), ISSN
  \bibinfo{issn}{1050-2947}.

\bibitem[{\citenamefont{Carusotto and Ciuti}(2013)}]{Carusotto2013}
\bibinfo{author}{\bibfnamefont{I.}~\bibnamefont{Carusotto}} \bibnamefont{and}
  \bibinfo{author}{\bibfnamefont{C.}~\bibnamefont{Ciuti}},
  \bibinfo{journal}{Reviews of Modern Physics} \textbf{\bibinfo{volume}{85}},
  \bibinfo{pages}{299} (\bibinfo{year}{2013}), ISSN \bibinfo{issn}{0034-6861}.

\bibitem[{\citenamefont{{Le Boit\'{e}} et~al.}(2014)\citenamefont{{Le
  Boit\'{e}}, Orso, and Ciuti}}]{LeBoite2014}
\bibinfo{author}{\bibfnamefont{A.}~\bibnamefont{{Le Boit\'{e}}}},
  \bibinfo{author}{\bibfnamefont{G.}~\bibnamefont{Orso}}, \bibnamefont{and}
  \bibinfo{author}{\bibfnamefont{C.}~\bibnamefont{Ciuti}},
  \bibinfo{journal}{Physical Review A} \textbf{\bibinfo{volume}{90}},
  \bibinfo{pages}{063821} (\bibinfo{year}{2014}), ISSN
  \bibinfo{issn}{1050-2947}.

\bibitem[{\citenamefont{Koch et~al.}(2010)\citenamefont{Koch, Houck, Hur, and
  Girvin}}]{Koch2010a}
\bibinfo{author}{\bibfnamefont{J.}~\bibnamefont{Koch}},
  \bibinfo{author}{\bibfnamefont{A.}~\bibnamefont{Houck}},
  \bibinfo{author}{\bibfnamefont{K.~L.} \bibnamefont{Hur}}, \bibnamefont{and}
  \bibinfo{author}{\bibfnamefont{S.}~\bibnamefont{Girvin}},
  \bibinfo{journal}{Physical Review A} \textbf{\bibinfo{volume}{82}}
  (\bibinfo{year}{2010}), ISSN \bibinfo{issn}{1050-2947}.

\bibitem[{\citenamefont{Kolovsky}(2011)}]{Kolovsky2011}
\bibinfo{author}{\bibfnamefont{A.~R.} \bibnamefont{Kolovsky}},
  \bibinfo{journal}{EPL (Europhysics Letters)} \textbf{\bibinfo{volume}{93}},
  \bibinfo{pages}{20003} (\bibinfo{year}{2011}), ISSN
  \bibinfo{issn}{0295-5075}.

\bibitem[{\citenamefont{Umucalilar and Carusotto}(2011)}]{Umucalilar2011}
\bibinfo{author}{\bibfnamefont{R.~O.} \bibnamefont{Umucalilar}}
  \bibnamefont{and} \bibinfo{author}{\bibfnamefont{I.}~\bibnamefont{Carusotto}}
  (\bibinfo{year}{2011}), \eprint{1104.4071}.

\bibitem[{\citenamefont{Oktel et~al.}(2007)\citenamefont{Oktel, Niţă, and
  Tanatar}}]{Oktel2007}
\bibinfo{author}{\bibfnamefont{M.}~\bibnamefont{Oktel}},
  \bibinfo{author}{\bibfnamefont{M.}~\bibnamefont{Niţă}}, \bibnamefont{and}
  \bibinfo{author}{\bibfnamefont{B.}~\bibnamefont{Tanatar}},
  \bibinfo{journal}{Physical Review B} \textbf{\bibinfo{volume}{75}}
  (\bibinfo{year}{2007}), ISSN \bibinfo{issn}{1098-0121}.

\bibitem[{\citenamefont{Dean et~al.}(2013)\citenamefont{Dean, Wang, Maher,
  Forsythe, Ghahari, Gao, Katoch, Ishigami, Moon, Koshino et~al.}}]{Dean2013}
\bibinfo{author}{\bibfnamefont{C.~R.} \bibnamefont{Dean}},
  \bibinfo{author}{\bibfnamefont{L.}~\bibnamefont{Wang}},
  \bibinfo{author}{\bibfnamefont{P.}~\bibnamefont{Maher}},
  \bibinfo{author}{\bibfnamefont{C.}~\bibnamefont{Forsythe}},
  \bibinfo{author}{\bibfnamefont{F.}~\bibnamefont{Ghahari}},
  \bibinfo{author}{\bibfnamefont{Y.}~\bibnamefont{Gao}},
  \bibinfo{author}{\bibfnamefont{J.}~\bibnamefont{Katoch}},
  \bibinfo{author}{\bibfnamefont{M.}~\bibnamefont{Ishigami}},
  \bibinfo{author}{\bibfnamefont{P.}~\bibnamefont{Moon}},
  \bibinfo{author}{\bibfnamefont{M.}~\bibnamefont{Koshino}},
  \bibnamefont{et~al.}, \bibinfo{journal}{Nature}
  \textbf{\bibinfo{volume}{497}}, \bibinfo{pages}{598} (\bibinfo{year}{2013}),
  ISSN \bibinfo{issn}{1476-4687}.

\bibitem[{\citenamefont{Schl\"{o}sser et~al.}(1996)\citenamefont{Schl\"{o}sser,
  Ensslin, Kotthaus, and Holland}}]{Schlosser1996}
\bibinfo{author}{\bibfnamefont{T.}~\bibnamefont{Schl\"{o}sser}},
  \bibinfo{author}{\bibfnamefont{K.}~\bibnamefont{Ensslin}},
  \bibinfo{author}{\bibfnamefont{J.~P.} \bibnamefont{Kotthaus}},
  \bibnamefont{and} \bibinfo{author}{\bibfnamefont{M.}~\bibnamefont{Holland}},
  \bibinfo{journal}{Europhysics Letters (EPL)} \textbf{\bibinfo{volume}{33}},
  \bibinfo{pages}{683} (\bibinfo{year}{1996}), ISSN \bibinfo{issn}{0295-5075}.

\bibitem[{\citenamefont{Kuhl and St\"{o}ckmann}(1998)}]{Kuhl1998}
\bibinfo{author}{\bibfnamefont{U.}~\bibnamefont{Kuhl}} \bibnamefont{and}
  \bibinfo{author}{\bibfnamefont{H.-J.} \bibnamefont{St\"{o}ckmann}},
  \bibinfo{journal}{Physical Review Letters} \textbf{\bibinfo{volume}{80}},
  \bibinfo{pages}{3232} (\bibinfo{year}{1998}), ISSN \bibinfo{issn}{0031-9007}.

\bibitem[{\citenamefont{Hofstadter}(1976)}]{Hofstadter1976a}
\bibinfo{author}{\bibfnamefont{D.}~\bibnamefont{Hofstadter}},
  \bibinfo{journal}{Physical Review B} \textbf{\bibinfo{volume}{14}},
  \bibinfo{pages}{2239} (\bibinfo{year}{1976}), ISSN \bibinfo{issn}{0556-2805}.

\bibitem[{\citenamefont{Ruostekoski et~al.}(2002)\citenamefont{Ruostekoski,
  Dunne, and Javanainen}}]{Ruostekoski2002}
\bibinfo{author}{\bibfnamefont{J.}~\bibnamefont{Ruostekoski}},
  \bibinfo{author}{\bibfnamefont{G.~V.} \bibnamefont{Dunne}}, \bibnamefont{and}
  \bibinfo{author}{\bibfnamefont{J.}~\bibnamefont{Javanainen}},
  \bibinfo{journal}{Physical Review Letters} \textbf{\bibinfo{volume}{88}},
  \bibinfo{pages}{180401} (\bibinfo{year}{2002}), ISSN
  \bibinfo{issn}{0031-9007}.

\bibitem[{\citenamefont{Jaksch and Zoller}(2003)}]{Jaksch2003}
\bibinfo{author}{\bibfnamefont{D.}~\bibnamefont{Jaksch}} \bibnamefont{and}
  \bibinfo{author}{\bibfnamefont{P.}~\bibnamefont{Zoller}},
  \bibinfo{journal}{New Journal of Physics} \textbf{\bibinfo{volume}{5}},
  \bibinfo{pages}{56} (\bibinfo{year}{2003}), ISSN \bibinfo{issn}{1367-2630}.

\bibitem[{\citenamefont{Mueller}(2004)}]{Mueller2004}
\bibinfo{author}{\bibfnamefont{E.~J.} \bibnamefont{Mueller}},
  \bibinfo{journal}{Physical Review A - Atomic, Molecular, and Optical Physics}
  \textbf{\bibinfo{volume}{70}}, \bibinfo{pages}{1} (\bibinfo{year}{2004}),
  ISSN \bibinfo{issn}{10502947}, \eprint{0404306}.

\bibitem[{\citenamefont{Spielman}(2009)}]{Spielman2009}
\bibinfo{author}{\bibfnamefont{I.~B.} \bibnamefont{Spielman}},
  \bibinfo{journal}{Physical Review A - Atomic, Molecular, and Optical Physics}
  \textbf{\bibinfo{volume}{79}}, \bibinfo{pages}{1} (\bibinfo{year}{2009}),
  ISSN \bibinfo{issn}{10502947}, \eprint{0905.2436}.

\bibitem[{\citenamefont{Murray et~al.}(2009)\citenamefont{Murray, \"{O}hberg,
  Gomila, and Barnett}}]{Murray2009}
\bibinfo{author}{\bibfnamefont{D.~R.} \bibnamefont{Murray}},
  \bibinfo{author}{\bibfnamefont{P.}~\bibnamefont{\"{O}hberg}},
  \bibinfo{author}{\bibfnamefont{D.}~\bibnamefont{Gomila}}, \bibnamefont{and}
  \bibinfo{author}{\bibfnamefont{S.~M.} \bibnamefont{Barnett}},
  \bibinfo{journal}{Physical Review A - Atomic, Molecular, and Optical Physics}
  \textbf{\bibinfo{volume}{79}}, \bibinfo{pages}{1} (\bibinfo{year}{2009}),
  ISSN \bibinfo{issn}{10502947}.

\bibitem[{\citenamefont{Lin et~al.}(2009)\citenamefont{Lin, Compton,
  Jim\'{e}nez-Garc\'{\i}a, Porto, and Spielman}}]{Lin2009}
\bibinfo{author}{\bibfnamefont{Y.-J.} \bibnamefont{Lin}},
  \bibinfo{author}{\bibfnamefont{R.~L.} \bibnamefont{Compton}},
  \bibinfo{author}{\bibfnamefont{K.}~\bibnamefont{Jim\'{e}nez-Garc\'{\i}a}},
  \bibinfo{author}{\bibfnamefont{J.~V.} \bibnamefont{Porto}}, \bibnamefont{and}
  \bibinfo{author}{\bibfnamefont{I.~B.} \bibnamefont{Spielman}},
  \bibinfo{journal}{Nature} \textbf{\bibinfo{volume}{462}},
  \bibinfo{pages}{628} (\bibinfo{year}{2009}), ISSN \bibinfo{issn}{1476-4687}.

\bibitem[{\citenamefont{Madison et~al.}(2000)\citenamefont{Madison, Chevy, and
  Wohlleben}}]{Madison2000}
\bibinfo{author}{\bibfnamefont{K.}~\bibnamefont{Madison}},
  \bibinfo{author}{\bibfnamefont{F.}~\bibnamefont{Chevy}}, \bibnamefont{and}
  \bibinfo{author}{\bibfnamefont{W.}~\bibnamefont{Wohlleben}},
  \bibinfo{journal}{Physical review letters} \textbf{\bibinfo{volume}{84}},
  \bibinfo{pages}{806} (\bibinfo{year}{2000}), ISSN \bibinfo{issn}{1094-5695},
  \eprint{9912015}.

\bibitem[{\citenamefont{Madison et~al.}(2001)\citenamefont{Madison, Chevy,
  Bretin, and Dalibard}}]{Madison2001}
\bibinfo{author}{\bibfnamefont{K.~W.} \bibnamefont{Madison}},
  \bibinfo{author}{\bibfnamefont{F.}~\bibnamefont{Chevy}},
  \bibinfo{author}{\bibfnamefont{V.}~\bibnamefont{Bretin}}, \bibnamefont{and}
  \bibinfo{author}{\bibfnamefont{J.}~\bibnamefont{Dalibard}},
  \bibinfo{journal}{Physical Review Letters} \textbf{\bibinfo{volume}{86}},
  \bibinfo{pages}{4443} (\bibinfo{year}{2001}), ISSN \bibinfo{issn}{00319007},
  \eprint{0101051}.

\bibitem[{\citenamefont{Hodby et~al.}(2001)\citenamefont{Hodby, Hechenblaikner,
  Hopkins, Marag\`{o}, and Foot}}]{Hodby2001}
\bibinfo{author}{\bibfnamefont{E.}~\bibnamefont{Hodby}},
  \bibinfo{author}{\bibfnamefont{G.}~\bibnamefont{Hechenblaikner}},
  \bibinfo{author}{\bibfnamefont{S.~A.} \bibnamefont{Hopkins}},
  \bibinfo{author}{\bibfnamefont{O.~M.} \bibnamefont{Marag\`{o}}},
  \bibnamefont{and} \bibinfo{author}{\bibfnamefont{C.~J.} \bibnamefont{Foot}},
  \bibinfo{journal}{Physical Review Letters} \textbf{\bibinfo{volume}{88}},
  \bibinfo{pages}{010405} (\bibinfo{year}{2001}), ISSN
  \bibinfo{issn}{0031-9007}.

\bibitem[{\citenamefont{Cooper and Wilkin}(1999)}]{Cooper1999}
\bibinfo{author}{\bibfnamefont{N.}~\bibnamefont{Cooper}} \bibnamefont{and}
  \bibinfo{author}{\bibfnamefont{N.}~\bibnamefont{Wilkin}},
  \bibinfo{journal}{Physical Review B} \textbf{\bibinfo{volume}{60}},
  \bibinfo{pages}{R16279} (\bibinfo{year}{1999}), ISSN
  \bibinfo{issn}{0163-1829}.

\bibitem[{\citenamefont{Wilkin and Gunn}(2000)}]{Wilkin2000}
\bibinfo{author}{\bibfnamefont{N.~K.} \bibnamefont{Wilkin}} \bibnamefont{and}
  \bibinfo{author}{\bibfnamefont{J.~M.~F.} \bibnamefont{Gunn}},
  \bibinfo{journal}{Physical Review Letters} \textbf{\bibinfo{volume}{84}},
  \bibinfo{pages}{6} (\bibinfo{year}{2000}), ISSN \bibinfo{issn}{0031-9007}.

\bibitem[{\citenamefont{Cooper et~al.}(2001)\citenamefont{Cooper, Wilkin, and
  Gunn}}]{Cooper2001}
\bibinfo{author}{\bibfnamefont{N.}~\bibnamefont{Cooper}},
  \bibinfo{author}{\bibfnamefont{N.}~\bibnamefont{Wilkin}}, \bibnamefont{and}
  \bibinfo{author}{\bibfnamefont{J.}~\bibnamefont{Gunn}},
  \bibinfo{journal}{Physical Review Letters} \textbf{\bibinfo{volume}{87}},
  \bibinfo{pages}{120405} (\bibinfo{year}{2001}), ISSN
  \bibinfo{issn}{0031-9007}.

\bibitem[{\citenamefont{Bretin et~al.}(2004)\citenamefont{Bretin, Stock,
  Seurin, and Dalibard}}]{Bretin2004}
\bibinfo{author}{\bibfnamefont{V.}~\bibnamefont{Bretin}},
  \bibinfo{author}{\bibfnamefont{S.}~\bibnamefont{Stock}},
  \bibinfo{author}{\bibfnamefont{Y.}~\bibnamefont{Seurin}}, \bibnamefont{and}
  \bibinfo{author}{\bibfnamefont{J.}~\bibnamefont{Dalibard}},
  \bibinfo{journal}{Physical Review Letters} \textbf{\bibinfo{volume}{92}},
  \bibinfo{pages}{050403} (\bibinfo{year}{2004}), ISSN
  \bibinfo{issn}{0031-9007}.

\bibitem[{\citenamefont{Schweikhard et~al.}(2004)\citenamefont{Schweikhard,
  Coddington, Engels, Mogendorff, and Cornell}}]{Schweikhard2004}
\bibinfo{author}{\bibfnamefont{V.}~\bibnamefont{Schweikhard}},
  \bibinfo{author}{\bibfnamefont{I.}~\bibnamefont{Coddington}},
  \bibinfo{author}{\bibfnamefont{P.}~\bibnamefont{Engels}},
  \bibinfo{author}{\bibfnamefont{V.~P.} \bibnamefont{Mogendorff}},
  \bibnamefont{and} \bibinfo{author}{\bibfnamefont{E.~A.}
  \bibnamefont{Cornell}}, \bibinfo{journal}{Physical Review Letters}
  \textbf{\bibinfo{volume}{92}}, \bibinfo{pages}{040404}
  (\bibinfo{year}{2004}), ISSN \bibinfo{issn}{0031-9007}.

\bibitem[{\citenamefont{Baranov et~al.}(2005)\citenamefont{Baranov, Osterloh,
  and Lewenstein}}]{Baranov2005}
\bibinfo{author}{\bibfnamefont{M.~A.} \bibnamefont{Baranov}},
  \bibinfo{author}{\bibfnamefont{K.}~\bibnamefont{Osterloh}}, \bibnamefont{and}
  \bibinfo{author}{\bibfnamefont{M.}~\bibnamefont{Lewenstein}},
  \bibinfo{journal}{Physical Review Letters} \textbf{\bibinfo{volume}{94}},
  \bibinfo{pages}{070404} (\bibinfo{year}{2005}), ISSN
  \bibinfo{issn}{0031-9007}.

\end{thebibliography}
\end{document}